\documentclass[published]{JHEP}

\usepackage{amssymb}

\def\m{\mu}
\def\n{\nu}
\def\a{\alpha}
\def\lb{w}
\def\bbox{\nabla^2}
\def\pt{{\tilde p}}
\def\mt{{\tilde m}}

\title{Induced brane gravity: realizations and limitations}

\author{Elias Kiritsis, Nikolaos Tetradis and Theodore N. Tomaras\\
Department of Physics, University of Crete, and\\ FO.R.T.H., 71003
Heraklion, Greece\\ E-mail: \email{kiritsis@physics.uoc.gr},
\email{tetradis@physics.uoc.gr}, \email{tomaras@physics.uoc.gr}}

\abstract{The induced 4d gravity on a brane world is analyzed.
The case of a thick brane is considered and the unexpected
appearance of a new threshold scale, much larger than the
thickness scale is found. In cases of phenomenological interest,
this new length scale turns out to be in the submillimeter range.
The effect of $R^2$ corrections, both in the bulk and on the
brane, is also studied. It is shown that they introduce new
threshold scales and may induce drastic modifications to the
leading behavior. A concrete string/D-brane realization of the
induced gravity scenario is also presented.}

\keywords{Extra Large Dimensions, p-branes, D-branes, Superstring Vacua}

\received{June 26, 2001}

\accepted{August 06, 2001}

\JHEP{08(2001)012}

\begin{document}

\section{Introduction}

We consider a 3-brane in a higher-dimensional bulk as a model of
our Universe. Gravity lives in the bulk, while the Standard Model
particles are confined on the brane. Quantum loops of the latter
will induce a four-dimensional Einstein term on the brane, which,
as was pointed out in~\cite{dgp} for the case of a zero-thickness
brane, can have surprising effects on the effective gravitational
interaction on it. For a five-dimensional bulk, the relevant part
of the action is parametrized as
\begin{equation}
S = M^3\int d^5x\, \sqrt{-g}R+M^3r_c\int d^4 x\sqrt{-\hat g}\hat
R\,,
\label{00}
\end{equation}
where ${\hat g}_{\alpha \beta}$, with $\alpha,\beta=0,1,2,3$, is
the induced metric on the 3-brane. When the fifth dimension is
non compact, the propagator of the graviton has the form $\sim
M^{-3} (p + r_c p^2 /2)^{-1}$, where $p$ is the magnitude of its
four-momentum along the brane. At length scales $l \sim p^{-1}
\gg r_c$ gravity behaves as in five dimensions, while in the
opposite limit $l \sim p^{-1} \ll r_c$ it is effectively four
dimensional.

It was further pointed out in~\cite{dgkn} that, upon
compactification of the fifth dimension on a circle of radius $R$
with $R\ll r_c$, the gravitational interaction is four
dimensional at all scales.\footnote{This may happen also without
compactification, see~\cite{dvv}.} This can be understood from
\pagebreak[3]the fact that gravity is four dimensional at length scales
smaller than $r_c$, while the compact circle makes spacetime
effectively four dimensional at length scales larger than $R$ as
well. There is an infinite discrete spectrum of graviton modes,
 with couplings to the brane fields that are
suppressed compared to the usual compactification scenario. This
modifies the appropriate experimental limits coming from solar
system motions as well as energy loss in stars and supernovae,
and allows values as exotic as $R\simeq 10^{-4} r_c \simeq
10^{16}\,\mathrm{m}$~\cite{dgkn}.

The purpose of this paper is to analyse the consistency and stability
of this scenario as well as its realization in string theory. There
are several aspects that merit investigation:

a) The idealization of the brane as infinitely thin may not be a
good approximation.\footnote{The effect of the brane thickness on
gravity on the brane has also been studied in~\cite{greg}.} Any
theory in which branes are intrinsic dynamical structures rather
than external sources predicts a non-zero value for the brane
thickness $\lb$. In such a theory there are (at least) two
arbitrary parameters: $M$ and $R$. The parameter $r_c$ is
determined by one-loop corrections to the Einstein term from
3-brane fields, while $\lb$ is the thickness of the solitonic
brane. We will evaluate $r_c$ in the context of a string model
later, and we expect $\lb$ to be of order $1/M$.\footnote{This is
not always necessary. We will see a different case in
section~\ref{sec:6}.} However, for most of this work we will
consider $r_c$ and $\lb$ as independent length scales that can be
varied arbitrarily. Furthermore, we will assume that $\lb\ll
R,r_c$. In the  limit $\lb\rightarrow 0$ we should recover the
results of~\cite{dgkn}. Eventually, we would like to take
$\lb^{-1}$ to be the ``near-new-physics'' scale of $\sim
10\,\mathrm{TeV}$.

We are interested in physics at energies small compared to the
thickness scale $\lb^{-1}$. At energies much larger than
$\lb^{-1}$, the theory may be strongly coupled, or even not
describable by field theory (as it may happen in string-theory
examples). In section~\ref{sec:2} we study the thick-brane
situation with a non-compact fifth dimension. We find that the
finite brane thickness produces an important modification to the
$\lb=0$ story: there is a new energy threshold appearing at
$E_b=1/\sqrt{\lb r_c}$. For energies $E_b\gg E \gg 1/r_c$ physics
is four dimensional. However, for $E\gg E_b$ the behavior depends
strongly on the exact location on the brane along the fifth
dimension. In particular, the gravitational potential between
points at a distance $r \simeq 1/E$ along the brane becomes five
dimensional at the center of the brane.

At energies much smaller than the scale set by the brane
thickness the detailed behavior across the brane cannot be
resolved. For this reason, the physical observables involve
averages along the fifth dimension. We study the behavior of such
observables in general terms. We show that, at energies below the
new threshold, their behavior deviates from four dimensional.
Moreover, the deviations are not universal for all sources
localized on the brane. For an observer who views such sources as
fundamental particles, this leads to violations of the
equivalence principle (This was observed independently
in~\cite{ru}).

In section~\ref{sec:3} we study the case of a compact
fifth dimension and analyse the graviton KK modes.
Those with masses $m \ll E_b$ are almost constant across
the brane and very similar to the
ones studied in~\cite{dgkn}. However, the modes with $m \gg E_b$
oscillate strongly between the two edges of the brane.

The compactified scenario is the only one consistent with the
observed four-dimensional nature of gravity at distances larger
than the solar system. For the allowed values of $\lb$, $r_c$ and
$R$, the new threshold appears at observable distances of the
submillimeter range.

b) The emergence of a new energy scale that is not apparent in
the action~(\ref{00}) raises the question of the importance of
the terms omitted in it. The four-dimensional Einstein term is
the first correction to the bulk action. The next corrections in
a derivative expansion of the action involve the $R^2$ term in
the bulk and the ${\hat R}^2$ term on the brane. In
sections~\ref{sec:4} and~\ref{sec:5} we study the behavior of the
propagator in their presence\footnote{The graviton propagator in
the presence of $R^2$ terms in the RS context was considered
in~\cite{kor}.}  and
find that they also induce important modifications. The
appearance of new threshold scales is generic in these enlarged
theories.

We should mention at this point that, following~\cite{dgp}, we
assume the absence of cosmological constant terms both in the
bulk and on the brane. This assumption represents two
fine-tunings for which no justification is provided by the
derivative expansion.

c) In order to have a theory with a controlled ultraviolet
behavior where this scenario is realized, we consider in
section~\ref{sec:6} the low-energy limit of a string construction
in which the evaluation of the $R$ and $R^2$ terms is possible
both on the brane and in the bulk. We demonstrate the appearance
of a threshold scale at which the four-dimensional gravitational
behavior breaks down. Moreover, we can predict in precise terms
the new behavior.

Our conclusions are presented in the final section.

\section{The non-compact case}\label{sec:2}

We start with the 5d gravitational action
\begin{equation}
S_5=M^3\int d^5 x\,\sqrt{-g}\,R\,.\label{1}
\end{equation}\looseness=1
In addition, we consider a brane, described by its collective
coordinates $X^{\mu}(\xi)$ with $\mu=0,1,2,3,4$, and carrying
massless fluctuating fields. Their action is of the~form
\begin{equation}
S_4=T_4\int d^4\xi \sqrt{-\hat g} L_{\mathrm{matter}}
(\hat g, \phi_i)\,,\label{2}
\end{equation}
where $\hat g_{\alpha\beta}=g_{\mu\nu}\partial_{\alpha}X^{\mu}
\partial_{\beta}X^{\nu}$ is the induced metric,
and $\phi_i$ denote collectively the other massless fields
localized on the brane. We pick the static gauge
$X^{\alpha}=\xi^{\alpha}$ and neglect the classical fluctuations
of the brane ($X^4\equiv z= \mathrm{constant}$). \pagebreak[3]

The action in~(\ref{2}) assumes an infinitely thin brane. When
the brane has a non-zero thickness the brane fields have a finite
extent in the fifth direction transverse to the brane. Let us
denote the brane thickness by $\lb$. Apart from special
situations, we expect that $\lb \sim 1/M$. \pagebreak[3]

The quantum fluctuations of the brane fields renormalize the bulk
action and in particular generate a correction to the
five-dimensional Einstein term. Due to the unbroken
four-dimensional Poincar{\'e} invariance, this is expected to be
of the form
\begin{equation}
S_{\mathrm{quantum}}=M^3r_c\int d^4 x\,\sqrt{-\hat g}\,\hat R\,,
\label{3}
\end{equation}
where the induced metric in the absence of fluctuations is
$g_{\m\n}(x^{\a},z=0)$, and $\hat R$ is the four-dimensional
curvature scalar constructed out of the induced metric.

In order to study the case of a brane of non-zero thickness we
replace~(\ref{3}) by
\begin{equation}
S_{\mathrm{quantum}}=M^3r_c \, \Delta(z) \int d^4 x\,\sqrt{-\hat g}
\,\hat R\,,\label{3thick}
\end{equation}
where $\Delta(z)=1/\lb$ for $z\in [-\lb/2,\lb/2]$ and zero
otherwise. The detailed structure of the brane and the resulting
corrections to eq.~(\ref{3thick}) should be irrelevant, as long
as we are interested in distances much larger than $\lb$. We will
check to what extent this expectation is fulfilled by the
solutions.

We compute the graviton propagator from the action
$S_5+S_{\mathrm{quantum}}$, as given by eqs.~(\ref{2})
and~(\ref{3thick}). We ignore the tensor structure and solve
instead for the scalar propagator $G(x^\alpha,z)$ (since this
captures the essential behavior)
\begin{equation}
M^3 \left( \bbox_{5}+r_c \Delta(z)\bbox_4 \right) G(x^\alpha,z)=
\delta^{(4)}(x^\alpha)\delta(z)\,, \label{4}
\end{equation}
where $\bbox_n$ is the flat laplacian in n dimensions. Going to
momentum space for the four-dimensional part we obtain the
equivalent equation (we work in euclidean space with $p^0=-ip^5$)
\begin{equation}
M^3 \left( p^2-\partial_z^2+r_c \Delta(z)p^2 \right)
G(p^{\alpha},z)=\delta(z)\,,\label{5}
\end{equation}
where $p^2=p^2_5+p^2_1+p^2_2+p^2_3$. The solution is
\begin{equation}
G\left(p^\alpha,z \right)=A\,e^{-p|z|}\,,\qquad \mbox{for}\qquad
|z|> \frac{\lb}{2}\label{g1}
\end{equation}
and
\begin{equation}
G(p^\alpha,z)=B\,e^{\pt |z|}+C~e^{-\pt |z|}\,,\qquad \mbox{for}\qquad
|z|< \frac{\lb}{2}\,,\label{g2}
\end{equation}
with
\begin{eqnarray}
\pt &=& \sqrt{1+\frac{r_c}{\lb}}\,p\,,\nonumber\\
A &=& \frac{2\pt}{p + \pt} e^{\frac{1}{2} (p-\pt)\;\lb}\,C\,,\qquad
B=\frac{\pt-p}{\pt+p}\, e^{-\pt\;\lb}\,C\,,\nonumber\\
C&=&\frac{1}{2 M^3} \frac{p+\pt}{\pt\left(p+\pt+(p-\pt)e^{-\pt\;\lb}
\right)}\,.\label{cc}
\end{eqnarray}
In the limit $\lb\rightarrow 0$, $\Delta(z)\rightarrow \delta(z)$
we approach asymptotically the case of the infinitely thin brane
analysed already in~\cite{dgkn}. In this case $\pt\rightarrow
\infty$ and $\pt\;\lb\rightarrow 0$, so that
\begin{equation}
A_0=2C_0=2B_0=\frac{1}{M^3(2p+ r_cp^2)}\,.\label{zerothick}
\end{equation}
The propagator exhibits four-dimensional fall-off for $p\gg
r_c^{-1}$ and five-dimensional fall-off for $p\ll r_c^{-1}$.

Going back to the non-zero thickness propagator we find that at
the center of the brane ($z=0$) it is given by
\begin{equation}
G(p^\alpha,0)=B+C=\frac{1}{2 M^3}\,\frac{p+\pt}{\pt\left(p+\pt+
(p-\pt) e^{-\pt\;\lb}\right)}\left(\frac{\pt-p}{\pt+p}\,e^{-\pt\;\lb}
+1 \right).\label{prz0}
\end{equation}
For $\lb\rightarrow 0$ this expression reduces
to~(\ref{zerothick}). In the region $\pt\;\lb \ll 1$ the
propagator can be approximated very well by the $\lb=0$ one.
Thus, the previous analysis suffices. For $\pt\;\lb \gg 1$, which
implies $p \gg (r_c\lb)^{-1/2}\equiv E_b$, the exponentials
die-off and we~obtain
\begin{equation}
G(p^\alpha,0)\simeq \frac{1}{2 M^3}\frac{1}{\pt}= \frac{1}{2 M^3}
\sqrt{\frac{\lb}{r_c}}\,\frac{1}{p}= \frac{1}{2 M^3 E_b r_c}
\frac{1}{p}\,.
\end{equation}
The propagator again exhibits five-dimensional behavior with an
effective Planck constant $M(r_c/\lb)^{1/6}$. Furthermore, a more
thorough examination of the solution reveals that it has peculiar
properties: near the surface of the brane $G(p^{\alpha},z)$ is
exponentially suppressed by $\pt\lb/2$ when this is large. In
fact, $G(p^{\alpha},z)$ is significantly different from zero only
within a region $|z| \lesssim \pt^{-1}$ around the center of the
brane.

At energy scales much smaller than the scale set by the brane
thickness the detailed $z$-dependence of the sources across the
brane cannot be resolved. For this reason one must define
effective correlations by averaging both the location of the
source and the point of observation along the fifth dimension
when computing the propagator. In order to achieve this we
calculate the propagator with the $\delta$-function located at
some point in the interval $[-\lb/2,\lb/2]$. Then we average over
the location of this point using a constant weight. Finally we
average over the point of observation within the brane. A
straightforward but lengthy calculation gives
\begin{equation}
{\bar G}=\frac{1}{M^3r_c p^2}\left(1-\frac{1}{\frac{pr_c}{2}+
\frac{p}{E_b}\coth \left( \frac{p}{E_b}\right)}\right).\label{avprop}
\end{equation}
The averaged propagator displays behavior very similar to that of
eq.~(\ref{zerothick}) at all scales. For $p\ll E_b$ this is
expected, as the solutions are almost constant across the brane.
However, it is rather surprising for $p\gg E_b$. This result
raises serious doubts on the physical relevance of the new scale
$E_b$, as its observable effects seem to disappear.

In order to address this important question we consider the
generalization of~(\ref{5}) for a source $J(z)$
\begin{equation}
M^3 \left( p^2-\partial_z^2+r_c \Delta(z)p^2 \right)G(p^{\alpha},z)
=J(z)\,.\label{5g}
\end{equation}\looseness=1
This is the equation of motion for the gravitational field, which
we also denote by $G$. We assume that $J(z)$ is different from
zero only within the brane boundaries and its integral in the
region $[-\lb/2,\lb/2]$ is unity. By integrating both sides of
this equation with respect to $z$ in an interval
$[-\delta/2,\delta/2]$, with $\delta$ a bit bigger than $\lb$,
we~obtain
\begin{eqnarray}
M^3 \left(\left. \frac{\partial G}{\partial z}\right|_{-\delta/2}
- \left.\frac{\partial G}{\partial z}\right|_{\delta/2}\right)
+\lb M^3 p^2 \int_{-\delta/2}^{\delta/2}G(p,z)\frac{dz}{\lb}
+\nonumber\\
+\,r_c M^3 p^2 \int_{-\lb/2}^{\lb/2}G(p,z)\Delta(z)dz &=&1\,.
\label{nik}
\end{eqnarray}
The first term in the above expression is negligible for $p\gg
E_b$. For general smooth $\Delta(z)$ we expect a strong
exponential fall-off of $G$ with $z$, so that it becomes
negligible outside the boundaries of the brane. The second term
is always expected to be smaller than the third one for $\lb\ll
r_c$ and smooth $\Delta(z)$. The conclusion is~that
\begin{equation}
\int_{-\lb/2}^{\lb/2}G(p,z)\Delta(z)dz=\frac{1}{M^3 r_c p^2}
\label{general}
\end{equation}
for $p\gg E_b$, to a very good approximation. Thus we confirm the
physical picture implied by~(\ref{avprop}), and also generalize
it to arbitrary smooth brane profiles~$\Delta(z)$.

However, the physical quantity of interest is not the field
averaged over the brane profile $\Delta(z)$, but over the profile
$J'(z)$ of some probe. The distributions $\Delta(z)$ and $J(z)$
or $J'(z)$ are not identical, because $\Delta(z)$ results from
loop corrections involving a multitude of intermediate states,
each with a different profile. This fact has important
phenomenological consequences: only the propagator averaged over
$\Delta(z)$ displays a pure four-dimensional behavior for $p \gg
E_b$. Depending on the profile of the probe $J'(z)$ deviations
from four-dimensional $\sim 1/r$ fall-off are expected. These can
be substantial, as one may verify by evaluating the average of
the solution~(\ref{g1})--(\ref{cc}) for various smooth functions
$J'(z)$. Moreover, these deviations are not universal and, for an
observer that considers the probes as fundamental particles, lead
to violations of the equivalence principle. We
conclude\footnote{These conclusions do not depend on the shape of
the thick-brane profile, as an analysis of other (smooth)
profiles indicates.} that the new scale $E_b$ has important
phenomenological consequences. Similar arguments have been given
in~\cite{ru} for the question of charge universality when gauge
fields are localized on the brane through a mechanism similar to that for gravity.

The interpretation of these results must take into account the
consistency of the whole picture. For $\pt\;\lb \ll 1$ the
propagator is almost constant along the fifth dimension
throughout the brane. Thus, we are within the region of validity
of the low-energy description in terms of the action of
eqs.~(\ref{1}) and~(\ref{3thick}). For $\pt\;\lb \gg 1$ the main
characteristic of the solution is the very strong variation of
the propagator across the brane. Such behavior is associated with
the high-energy regime, which we have assumed that has been
integrated out in order to derive the low-energy action. This
action seems to become insufficient for a consistent description
at a scale $\sqrt{r_c \lb} \gg \lb$, contrary to our initial
expectation. Thus the theory displays peculiar non-decoupling
behavior.

In the following section we will see another manifestation of this
behavior in the case of a compact fifth dimension. There,
averaging the propagator corresponds to averaging strongly
oscillatory KK modes on the brane. Therefore, it must be performed
in a systematic fashion (through the use of the renormalization
group for example) in order to have physical significance.
Consequently the proper averaging depends on the (unknown) details
of the fundamental theory.

The moral of this section is that theories with induced gravity
on the brane exhibit a breakdown of predictability in the
gravitational sector at an energy scale hierarchically lower than
the fundamental one. In section~\ref{sec:6} we will describe an
example where the fundamental theory is controllable, and where
we will be able to discern the physics beyond the new threshold
scale.

\section{Compactification}\label{sec:3}

We compactify the fifth dimension on a circle of radius $R \gg
\lb$. The compactification does not change the short-distance
structure of the theory, while it turns the five-dimensional
long-distance behavior into a four-dimensional one.
In~\cite{dgkn} $R$ was taken a few orders of magnitude smaller
than $r_c$, so that gravity is four dimensional to a good
approximation at all astrophysical scales.

The calculation of the propagator is completely analogous to the
one in the previous section. The only difference is the use of
periodic boundary conditions. We give the result for the interior
of the brane, where
\begin{equation}
G(p^\alpha,z)=B~e^{\pt |z|}+C~e^{-\pt |z|}\,,\label{propnc}
\end{equation}
with
\begin{eqnarray}
C &=& \frac{1}{2 M^3\pt} \left[1+\frac{{p-\pt \coth\left(p
\frac{R-\lb}{2}\right)}} {p+\pt \coth\left(p \frac{R-\lb}{2}\right)}
e^{-\pt \lb} \right]^{-1},\nonumber\\
B &=& C-\frac{1}{2 M^3\pt}\,.\label{bnc}
\end{eqnarray}

We are also interested in the spectrum of the KK modes resulting
from this large-scale compactification. As before we ignore the
tensorial structure and consider the case of a scalar field
$\Phi(x^\alpha,z)$. The equation of motion for the KK modes is
(we revert to Minkowski space)
\begin{equation}
\left( \partial_z^2+m_n^2+r_c \Delta(z)m_n^2 \right) \phi_n(z)=0\,,
\label{31}
\end{equation}
where $\Phi(x^\alpha,z)=\sum_n \phi_n(z)\sigma_n(x^\alpha)$ and
the $\sigma_n(x^\alpha)$ satisfy the four-dimensional
Klein-Gordon equation $(\partial^\alpha
\partial_\alpha+m_n^2)\sigma_n=0$.

The solutions have the form (for simplicity we omit the index $n$
from $m_n$, $\phi_n$)
\begin{eqnarray}
\phi(z) &=& A\,e^{i\mt z}+B\, e^{-i\mt z}\,,\qquad \mbox{for}\qquad
|z|< \frac{\lb}{2}\,,\label{kk1}\\
\phi(z) &=& C\, e^{im z}+D~e^{-im z}\,,\qquad \mbox{for}\qquad
\frac{\lb}{2} < z < R- \frac{\lb}{2}\,,\label{kk2}
\end{eqnarray}
with
\begin{equation}
\mt=\sqrt{1+\frac{r_c}{\lb}}\,m\,.
\end{equation}
The periodicity condition permits solutions only with discrete
values of $m$. There are two classes of solutions, depending on
the reflection properties of~(\ref{kk2}) around the center of the
brane (at $z=0$). The symmetric class has $A=B$, while the
antisymmetric $A=-B$. The quantization conditions for the masses
$m$ are
\begin{equation}
\tan\left( m\frac{R-\lb}{2}\right) = - \left(\frac{\mt}{m}
\right)^{\pm 1} \tan\left(\frac{\mt \lb}{2} \right),\label{quant}
\end{equation}
with the $+(-)$ sign corresponding to the symmetric
(antisymmetric) class. An overall amplitude of the solutions is
determined through the normalization condition for the~field.

For gravitational sources located at the center of the brane,
such as the ones we are considering here, the antisymmetric
solutions do not contribute to the propagator, which is a
symmetric function of $z$. Thus, we concentrate on the symmetric
sector. By switching to Minkowski space it can be checked that
the propagator of eq.~(\ref{propnc}) has poles at the values
given by eq.~(\ref{quant}). In the limit $\lb \rightarrow 0$, the
quantization condition gives
\begin{equation}
\tan\left(\frac{mR}{2} \right) = - \frac{m r_c}{2}\,,\label{quants}
\end{equation}
in agreement with~\cite{dgkn}.

The behavior we observed in the previous section persists even
after the compactification. For $p \ll E_b$ the propagator is
almost constant across the brane and the results reproduce those
of the scenario with an infinitely thin brane. For $p \gg E_b$
(or $\pt\lb \gg 1$) the propagator is exponentially suppressed,
apart from a region $|z| \lesssim \pt^{-1}$ near the center of
the brane. The strong variation of the propagator indicates that
we are outside the region of validity of the low-energy
description. Corrections originating in the specific structure of
the brane become relevant not at a length scale $\sim \lb$ as
expected, but at the much larger length scale $E_b^{-1}=\sqrt{\lb
r_c}$.

The form of the KK modes confirms these conclusions: for $m \ll
E_b$ the KK solutions are almost constant across the brane, while
for $m \gg E_b$ they become strongly oscillatory. Averaging the
propagator across the brane corresponds to averaging the strong
fluctuations of the KK modes in the interior of the brane. The
correct form of this procedure will be dictated by the
fundamental theory.

As a final point we examine the suppression of the high KK modes
relative to the zero mode on the brane. This is quantified by the
ratio
\begin{eqnarray}
\frac{\left| \phi_0(0) \right|^2}{\left| \phi_n(0) \right|^2}
&=& \cos^2 \frac{\mt\lb}{2}\left[\frac{1}{2}\left(1-\frac{\lb}{R}
\right) \left( 1 + \frac{\mt^2}{m^2} \tan^2 \frac{\mt\lb}{2} \right)
-\frac{1}{m R} \, \frac{\mt}{m} \tan \frac{\mt \lb}{2}\right]+
\nonumber\\
&&+\, \frac{\lb}{2R} + \frac{1}{2 \mt R} \sin \mt \lb\,,
\label{suppress}
\end{eqnarray}
with $n > 0$. For $m \ll E_b$ one recovers the suppression factor
$\sim m^2$ of~\cite{dgkn}, with the masses $m$ being approximately
$(2 n +1)\pi/R$ and $n$ a large integer. For $m \gg E_b$ the
solution of eqs.~(\ref{quant}) and~(\ref{suppress}) indicates
that there are unsuppressed modes with masses that are
approximately $(2 n +1) \pi E_b$. However, averaging these
strongly oscillatory modes across the brane generates a
suppression factor $\sim m$ for $\bar \phi$. As a result the
averaged KK modes are still suppressed by $\sim m^2$ relative to
the zero mode.

The compactified scenario is the only one that is
phenomenologically viable~\cite{dgkn}. In particular, $R$ must be
chosen much smaller than $r_c$, so that the gravitational
interaction has four-dimensional behavior to a good approximation
even at distances larger than the solar system. Reproducing
correctly Newton's constant requires $M^3 r_c = M^2_{Pl}$.
Assuming that the brane thickness is of order $M$ (the natural
choice) implies that the new threshold scale is
\begin{equation}
\frac{E_b^{-1}}{\mathrm{mm}} \sim 2.4 \;\left( \frac{M}{\mathrm{TeV}}
\right)^{-2} = 5.6 \times 10^4\;\left(\frac{r_c}{10^{20}{\mathrm{m}}}
\right)^{2/3}.\label{thresh}
\end{equation}

The parameter ranges allowed by observational constraints on
planetary motion are: a) $R\gtrsim 10^{16}\,\mathrm{m}$ with
$R/r_c=10^{-4}$, and b) $R$ smaller than the Earth radius and $M$
in the TeV range. For the first range, the new threshold appears
at a distance of several meters, a result that is in conflict
with experiment. For the second range, if $M\sim
10\,\mathrm{TeV}$, then $E_b^{-1}\sim 20\,\mbox{$\mu$m}$.

\section{The $R^2$ terms}\label{sec:4}

In this section we consider the higher-order corrections in a
derivative expansion of the action~({\ref{00}). Among these, the
terms relevant for the graviton propagator are the $R^2$ terms in
the bulk and on the brane. It is known that out of the three
independent tensor structures, a linear combination (the
Gauss-Bonnet term) does not contribute to the propagator. Our
discussion thus, applies to the opposite case. As in the previous
sections, we ignore the tensor structure and solve for the scalar
propagator in the presence of four-derivative terms in the
action.  Moreover, we ignore related factors of order one that
could appear in the equations of motion when they are derived
through variation of the action. We assume the bulk to be five
dimensional.

First we consider the corrections localized on the 3-brane, which
can be studied in the thin-brane limit. Let us parametrize them as
\begin{equation}
L_4=\sqrt{-\hat g} \left(M^3r_c\, {\hat R}+\lambda_2\,{\hat R}^2
\right).\label{a1}
\end{equation}
The effective coefficients $r_c$, $\lambda_2$ include quantum
corrections from the brane fields.

The propagator has the general form
\begin{equation}
G\sim \frac{1}{2p+r_cp^2+\lambda_2~p^{4}/M^3}\,.\label{a2}
\end{equation}
We are considering four-momenta $p\ll M$. In the regime $pr_c\ll
1$, we obtain a five-dimensional behavior provided that
\begin{equation}
\lambda_2\ll (Mr_c)^3\,.\label{a3}
\end{equation}
On the other hand, for $pr_c\gg 1$ we obtain a four-dimensional
behavior provided that $Mr_c\geq 1$ (in particular, we are
interested in $M r_c \gg 1$) and
\begin{equation}
\frac{\lambda_2}{Mr_c}\left(\frac{p}{M}\right)^{2}\ll 1\,.\label{a4}
\end{equation}
If
\begin{equation}
\lambda_2 < Mr_c \label{a5}
\end{equation}
the four-dimensional behavior extends up to the fundamental scale
M. However, if $\lambda_2\sim (Mr_c)^a$ with $1\leq a\leq 3$, the
four-dimensional behavior persists up to a scale $E'_b\sim
M~(Mr_c)^{(1-a)/2}$. This can be well below $M$ (since by
assumption $Mr_c\gg 1$). In particular, $\lambda_2$ can behave
either as $Mr_c$ or $(Mr_c)^2$. In the first case the
four-dimensional behavior extends up to the scale $M$, while in
the second there is a transition at $E'_b\sim \sqrt{M/r_c}$.

The $R^2$ corrections in the bulk are more delicate. A thin-brane
approach, treating the bulk $R^2$ terms perturbatively, runs into
problems with divergences, because squares of the
$\delta$-function appear. Therefore, it is obvious that a
short-distance regularization is needed. The simplest way to
implement one is to consider a thick brane, as in the previous
sections.

Thus, we need to solve the equation
\begin{equation}
\left[p^2-\partial^2_z+L^2(p^2-\partial_z^2)^2+\Delta(z)r_cp^2\right]
G=\delta(z)\,, \label{r2}
\end{equation}
where the length scale $L$ controls the strength of the $R^2$
term and $\Delta(z)$ is defined in section~\ref{sec:2}. For
$|z|>\lb/2$ the solution remains as before. For $|z|<\lb/2$ it
has exactly the same form as before, but $\tilde p$ is now given
by
\begin{equation}
\tilde p=\sqrt{p^2+\frac{1\pm\sqrt{1-4\frac{r_c}{\lb}L^2p^2}}
{2L^2}}\,.\label{LL}
\end{equation}
We are interested in solutions that are finite in the limit
$L\rightarrow 0$. This rules out the solution with the positive
sign, which scales $\sim 1/L$ in this limit. The presence of a
new scale $\sim \sqrt{\lb/r_c|L^2|}$ is apparent in this
expression ($L^2$ can be positive or negative). For
$p^2|L^2|{r_c/ \lb}\ll 1$ we obtain
\begin{equation}
\tilde p=p\sqrt{1+\frac{r_c}{\lb}+p^2L^2\frac{r_c^2}{\lb^2}+\cdots}
\end{equation}
and for the propagator
\begin{equation}
G\sim \frac{1}{2p+r_cp^2(1+\frac{r_c}{\lb}p^2L^2+\cdots)}\,.
\end{equation}

In the (natural) case $|L^2| \sim \lb^2$, the new scale is the
one found in the previous sections $\sim (\lb r_c)^{-1/2}$. For
$|L^2|$ smaller than $\lb^2$, it shifts to larger energies. We
see again that the detailed structure of the theory affects the
low-energy behavior drastically. A simple power-counting argument
would indicate that the $R^2$ correction is expected to be
negligible relative to the leading $R$ term at energies below
$1/L$. However, for $w \ll r_c$ the new energy scale appears in
this regime, and its exact value depends crucially on the
coefficient of the $R^2$ term. This implies that the effect of
higher terms in the derivative expansion is difficult to estimate
in theories of induced gravity on a brane living in a
higher-dimensional bulk. We will see similar examples in the
following section.

\section{More transverse dimensions}\label{sec:5}

For future applications we would like to discuss the case with
more than one large transverse dimensions~\cite{dg2}. We work in
the thin-brane limit. Starting from
\begin{equation}
S=M^{n+2}\int d^{n+4}x \sqrt{-g} ~R_{n+4}+M^{n+2}r_c^n\int d^4x
\sqrt{-\hat g}\,\hat R_4\label{a6}
\end{equation}
we obtain the propagator on the brane
\begin{equation}
G_n(p,0)\sim \frac{D_n(p,0)}{1+r_c^np^2D_n(p,0)}\,,\qquad
D_n(p,0)= \int \frac{d^n q}{p^2+q^2}\,.\label{prop1}
\end{equation}
$D_n(p,0)$  is UV divergent. In order to define it we introduce a
momentum cutoff $\Lambda$, which can be thought of as an
alternative to the finite-thickness regularization of the previous
section. Then
\begin{equation}
D_2(p,0)\sim \log\left[\frac{(\Lambda^2+p^2)}{p^2}\right],\qquad
D_4(p,0)\sim \Lambda^2- p^2\log\left[\frac{(\Lambda^2+p^2)}{p^2}
\right].
\end{equation}

For $n=4$ the propagator for momenta well below the cutoff is
\begin{equation}
G_4(p,0)\sim \frac{1}{p^2+ \frac{1}{r_c^4\Lambda^2}}\,.
\end{equation}
Thus, if $r_c\Lambda\gg 1$, for $\Lambda \gg p \gg (\Lambda
r_c^2)^{-1}$ we have four-dimensional gravity, while for $p\ll
(\Lambda r_c^2)^{-1}$ the gravitational force is screened.

Next we analyse the effect of bulk $R^2$ corrections in the
higher-dimensional case. The perturbation in the action is
parametrized by $M^{n+2}L^2\,R_{n+4}^2$. The solution for the
four-dimensional propagator is given by~(\ref{prop1}), where now
\begin{equation}
D_n(p,0)=\int d^nk \frac{1}{(k^2+p^2+L^2(k^2+p^2)^2)}\,.
\end{equation}
The appearance of the $R^2$ contribution in the propagator
regulates the UV divergence for $n<4$.

For $n=2$ we obtain\footnote{In the following we omit factors of
order one in the evaluation of $D_n(p,0)$. For this reason the
estimates for the various threshold scales are only approximate.}
\begin{equation}
D_2(p,0) \sim - \log\left[1+\frac{1}{p^2L^2}\right].
\end{equation}
For $pL\ll 1$ and $pr_c\ll 1$ we obtain a propagator $\sim
\log(p^2L^2)$. For $pL\ll 1$ but $pr_c\gg 1$ a four-dimensional
behavior is obtained with $G_N\sim (M^4r_c^2)^{-1}$. Finally, for
$pL\gg 1$ we still have a four-dimensional behavior with a
different Newton's constant, $G_N\sim (M^4(r_c^2-L^2))^{-1}$.

For $n=4$ $D_4$ is UV divergent. We introduce a cutoff and obtain
\begin{equation}
D_4(p,0)\sim p^2\log\left[\frac{(p^2+\Lambda^2)L^2+1}{p^2 L^2+1}
\frac{p^2}{p^2+\Lambda^2}\right] +\frac{1}{L^2}\log \left[
\frac{(p^2+\Lambda^2)L^2+1}{p^2 L^2+1}\right].
\end{equation}
As before, taking $\Lambda = \infty$ while keeping L fixed we
obtain a uniform four-dimensional behavior. Another natural
choice is that the cutoff scale determines the $R^2$ term:
$\Lambda L\sim 1$. In this case, we find the same behavior as in
the $L=0$ case: a massive graviton with mass $\sim 1/\Lambda
r_c^2$.

Clearly, there are more possibilities. We will always assume that
$L\Lambda\gg 1$ and $r_c\Lambda\gg 1$. The approximate forms of
$D_4(p,0)$ are
\begin{eqnarray}
pL &\ll & 1\,,\qquad D_4\sim \frac{2}{L^2}\log(L\Lambda)\,,
\nonumber\\[2pt]
pL &\gg & 1\,, \qquad \mbox{and} \qquad p\ll\Lambda\,,\qquad D_4\sim
\frac{1}{L^2} \log \frac{\Lambda^2}{p^2}\,,\nonumber\\[2pt]
pL &\gg & 1\,, \qquad \mbox{and}\qquad p\gg \Lambda\,,\qquad
D_4\sim \frac{\Lambda^4}{2L^2p^4}\,.
\end{eqnarray}
There are three possible orderings of the various scales:
\\
(i) $L/r_c^2\ll 1/r_c\ll 1/L\ll \Lambda$: when $pL\ll 1$
\begin{equation}
G_4\sim \frac{\log(L\Lambda)}{L^2+2p^2r_c^4\log(L\Lambda)}\,.
\label{aa1}
\end{equation}
When $p\ll L/r_c^2$ there is screening, while when $p\gg L/r_c^2$
we obtain $G_4\sim p^{-2}$. For $1/L \ll p \ll \Lambda$ we still
have $G_4\sim p^{-2}$, while for $p\gg\Lambda$
\begin{equation}
G_4\sim \frac{\Lambda^4}{(2L^2p^2+r_c^4\Lambda^4)p^2}\,.
\end{equation}
For $\Lambda \ll p\ll r_c^2\Lambda^2/L$ we still have
four-dimensional behavior, while for $p\gg r_c^2\Lambda^2/L$ we
obtain $G_4\sim \Lambda^4/(p^2L)^2$. To summarize, there is
screening for $p\ll L/r_c^2$, $1/p^{2}$ behavior for $L/r_c^2\ll
p\ll r_c^2\Lambda^2/L$ and $1/p^{4}$ behavior above.
\\
(ii) $1/L\ll 1/r_c\ll L/r_c^2\ll \Lambda$: for $p\ll 1/L$ there
is screening. For $L/r_c^2\gg p\gg 1/L$ we find $G_4\sim
\log[p^2/\Lambda^2]/L^2$. When  $\Lambda\gg p\gg L/r_c^2$,
$G_4\sim 1/p^2$. For $\Lambda^2r_c^2/L\gg p\gg \Lambda$ we still
have $1/p^2$ behavior which turns to $1/p^4$ for $p \gg
\Lambda^2r_c^2/L$.
\\
(iii) $1/L\ll 1/r_c\ll \Lambda\ll L/r_c^2$: the situation for
$p\ll \Lambda$ is as in case (ii). For $p\gg \Lambda$ we obtain
$1/p^4$ behavior.

\looseness=1
We conclude that on many occasions the $R^2$ terms can modify
gravity substantially in various regimes introducing new
threshold scales. The important observation is that these
modifications may occur in energy regions where the
higher-derivative terms such as $R^2$ are expected to be
negligible. For example, in the case (i) for $n=4$ discussed
above the lower energy scale $L/r_c^2$ is much smaller than
$1/L$. At such low energies the $R^2$ term should be completely
negligible relative to the $R$ term. However, the exact value of
the new scale depends on the coefficient $1/L$ of the $R^2$~term.

\looseness=1
There are also effects that have to do with the tensorial
structure of gravity and the fact that $R^2$ terms affect this
structure. These additional subtler effects eventually need to be
taken into account as well, even though we will not do this~here.

\section{D-brane realization}\label{sec:6}

String theory provides so far the unique example of a theory with
a controllable quantum gravity (below the Planck scale) as well
as ultraviolet-finite higher-dimensional theories
(see~\cite{string} for introductory reviews). Moreover, it has
provided us with calculable examples of brane solitons, namely
D-branes, which are the concrete realizations of the brane-world
idea. \pagebreak[3]

We will consider here type-I vacua with $N=2$ supersymmetry.
Realizations with less supersymmetry are possible but we will try
to keep the structure as simple as possible. The general
structure of such vacua is given by a collection of D9- and
D5-branes embedded in a ten-dimensional bulk~\cite{sagn,gp}. We
will consider two of the dimensions parallel to the  branes to be
compactified on a two-torus. The four extra dimensions are those
of a non-compact $\mathbb{R}^4/\mathbb{Z}_2$ orbifold with the
D5-branes located at the fixed point at the origin. Such vacua
are decompactifications of four-dimensional $N=2$ type-I vacua
with D5-branes when the K3 volume becomes infinite.

In such vacua there are eight infinite dimensions felt by the
closed string sector (gravity) and D9-branes. The D5-branes
(wrapped on $T^2$) have four infinite dimensions. Let $T,U$ be
the standard K{\"a}hler and complex structure moduli of $T^2$.
For two orthogonal circles of radii $R_{1,2}$ they are given by
$T=iR_1R_2$, $U=iR_1/R_2$. The following T-dualities are relevant:
\begin{eqnarray}
R_2 &\longrightarrow& \frac{1}{R_2} \Longrightarrow T_2
\longleftrightarrow U_2\,, \nonumber\\
R_1 &\longleftrightarrow& R_2 \Longrightarrow T_2\longrightarrow
T_2\,, \qquad U_2\longrightarrow \frac{1}{U_2}\,,\nonumber\\
R_1 &\longrightarrow& \frac{1}{R_1}\Longrightarrow
T_2\longleftrightarrow \frac{1}{U_2}\,.
\end{eqnarray}

The relevant gravitational terms in the low energy effective
action are
\begin{equation}
S=M^6\int d^8 x\, R_8 + M^6 \; L^2 \int d^8 x\, R^2_8+\cdots +
M^6r_c^4\int d^4x\,R_4 + \lambda_2\int d^4x\, R_4^2+\cdots\,.
\label{s1}
\end{equation}
The Einstein term in eight dimensions comes from the sphere
\begin{equation}
M^6=\frac{T_2}{g_s^2}M_s^6\,, \label{s2}
\end{equation}
where $M_s$ is the string scale, $g_s$ is the (ten-dimensional)
string coupling, $T_2$ the volume of the two torus, and we ignore
factors of order one.

We now focus on the $R_8^2$ term. In the case of the type-I
string in ten dimensions ($N = 4$ supersymmetry) there is an $R^2$
contribution coming from the disk. Moreover it is
known~\cite{review} that there are no further perturbative
corrections to its effective coupling. There are however
non-perturbative corrections to it coming from D5-brane
instantons~\cite{hmor}, that translate via heterotic/type-I
duality to NS5-instantons on the heterotic side and via
heterotic/type-II duality to fundamental string instantons on the
type-II side~\cite{hm,k4}. However, such corrections vanish in
eight non-compact dimensions (they also vanish in the case where
five dimensions are non compact, that we will discuss below). The
only other corrections to the $R_8^2$ terms may come from the
projected D9 spectrum. These  contributions can be obtained from
heterotic duals~\cite{kkpr} and can be shown to vanish in the
decompactification limit. Thus,
\begin{equation}
M^6\; L^2\sim \frac{T_2}{g_s}M_s^4\Longrightarrow L^2 \sim g_s\;
M_s^{-2}\,.\label{s3}
\end{equation}

The coefficient of the four-dimensional Einstein term can be
inferred from the calculations in~\cite{triality}. There, it was
pointed out that, in type-I vacua of the type discussed here,
there is a one-loop correction to the Einstein term. This matches
via heterotic/type-I duality to the universal contribution to
gauge couplings~\cite{kk} as well as to the one-loop K{\"a}hler
potential~\cite{anton}. The calculations were carried out in four
dimensions, but upon decompactication the contributions of the
9--5 and 5--5 sectors remain and give a purely four-dimensional
contribution (originally six dimensional, it becomes four
dimensional due to the torus compactification). We obtain
\begin{equation}
M^6r_c^4\sim \frac{M_s^2}{T_2}E_2(U,\bar U)= \frac{M_s^2}{T_2}
\sum_{(m,n)\in (Z,Z)-(0,0)}\frac{U_2^2}{|m+nU|^4}\,.\label{s4}
\end{equation}
$E_2$ is modular invariant and has the following asymptotic
expansion
\begin{equation}
E_2(U,\bar U)=2\zeta (4)U_2^2 + \frac{\pi\zeta (3)}{U_2} +
\mathcal{O}(e^{-\pi U_2})\,.\label{s5}
\end{equation}

Finally, the $R_4^2$ term comes from the one-loop corrections of
the 5--9 and the 5--5 states. The corrections are logarithmic and,
extending the results of~\cite{triality}, we obtain
\begin{equation}
\lambda_2\sim \log(T_2U_2|\eta(U)|^4)\,.\label{s6}
\end{equation}

We  will further compactify three out of the four dimensions of
the orbifold with radii of the order of the string scale. This
may bring several D5-branes at finite distance of the order of the
string scale, as there will be one at each fixed point of the
orbifold. Other configurations are also possible with D5-branes
being located at a single fixed point. In this case we do not
have local tadpole cancellation, and non-trivial one-loop
corrections to the tension can appear.

The effective action now becomes
\begin{equation}
S=M^3\int d^5 x\,R_5+M^3\; L^2\int d^5 x\,R^2_5+\cdots+M^3r_c\int
d^4x\,R_4+\lambda_2\int d^4x\,R_4^2+\cdots\,.\label{s7}
\end{equation}
with
\begin{eqnarray}
M^3 &=& \frac{T_2}{g_s^2}M_s^3\,,\qquad M^3\;L^2\sim \frac{T_2}{g_s}
M_s\,,\label{s8}\\
M^3r_c &\sim& \frac{M_s^2}{T_2}E_2(U,\bar U)
\end{eqnarray}
and $\lambda_2$ given in~(\ref{s6}). \pagebreak[3]

Thus we find $Mr_c\sim g_s^{4/3}T_2^{-5/3}E_2(U,\bar U)$, and
$M_sr_c\sim  E_2 g_s^2/T_2^2$. Since we must be working in a
weekly coupled description, which means that the volume should
always be greater or equal to the string length we must consider
the following inequivalent~cases:

(1) $R_{1,2}\gg 1$ in string units, which implies that $T_2 \gg
1$ and $E_2\sim \mathcal{O}(1)$. In this case $M^3r_c\sim 1/T_2$
and $\lambda_2\sim \log(T_2)$.

(2) $R_1\gg 1$, $R_2\sim \mathcal{O}(1)$. In this case $T_2\sim
U_2\gg 1$, $M^3r_c\sim E_2(T_2)/T_2\sim T_2$ and $\lambda_2\sim
T_2$.

(3) $R_1\gg 1$ and $R_2\ll 1$. In this case to go to a weakly
coupled description we must T-dualize once, and we will be
describing D4-branes with one transverse and one longitudinal
direction large. In this case $T_2\sim \mathcal{O}(1)$ and
$U_2\gg 1$. Upon T-duality $T_2\leftrightarrow U_2$ with $T_2\gg
1$ and $U_2\sim \mathcal{O}(1)$. Then $M^3r_c\sim
E_2(T_2)/U_2\sim T_2^2$ and $\lambda_2\sim T_2$.

(4) $R_1\sim R_2\sim \mathcal{O}(1)$. Then $M^3r_c\sim
\mathcal{O}(1)$ and $\lambda_2\sim \mathcal{O}(1)$.

(5) $R_{1,2}\ll 1$, so that $U_2\sim \mathcal{O}(1)$, $T_2\ll 1$.
In this case we will T-dualize twice and obtain D3-branes with
two large transverse dimensions. In the T-dual version
$M^3r_c\sim T_2\gg 1$ and $\lambda_2\sim \log(T_2)$.

From the various cases above only (3) can provide that $1/r_c$ is
much smaller both from the string scale and the five-dimensional
Planck scale. In particular $M_s r_c\sim g_s^2T_2 \gg 1$ and
$Mr_c\sim (g_sT_2)^{4/3}\gg 1$. Other  relevant scales in this
case are the compactification scale $M_c\sim M_s/\sqrt{T_2}$
which determines the decompactification of the two compact
dimensions, the 5-d Planck scale $M_{(5)}\sim
M_sg_s^{-2/3}T_2^{1/3}$, the eight-dimensional Planck scale
$M_{(8)}\sim M_s g_s^{-1/3}T_2^{1/6}$ as well as the
ten-dimensional Planck scale $M_{(10)}\sim M_s g_s^{-1/4}$.

We are in the perturbative regime with $g_s<1$ and $g_sT_2\gg 1$.
The ordering of the relevant scales is
\begin{equation}
\frac{1}{r_c} \ll M_c\ll M_s \ll M_{(10)}\ll M_{(8)}\ll M_{(5)}\,.
\end{equation}
At scales $E\ll 1/r_c$ gravity on the D4-brane is five
dimensional. For $1/r_c\ll E\ll M_c$ gravity on D4 is four
dimensional. In the regime $M_c\ll E\ll M_s$ the two dimensions
decompactify and physics (and gravity) in the bulk and on the
D4-brane is seven dimensional. For $M_s\ll E\ll M_{(10)}$ the
other three compact directions decompactify. Moreover stringy
effects become important. Thus, the bulk and brane physics is
that of ten-dimensional non-compact string theory. Finally, for
$E\gg M_{(10)}$ gravitational effects become strong and it is
believed that the physics is dominated by ten-dimensional
black-hole formation. Note that the $M_{(5)}$ and $M_{(8)}$
Planck scales are ``mirage'' scales since they do not correspond
to physical thresholds of the theory.

Let us now analyse  the $R^2$ corrections on the brane for case
(3). In~\cite{bbg} it was shown that $R^2$ corrections on
D-branes, unlike those in the bulk, do not come in the
Gauss-Bonnet combination and are thus contributing to the
graviton propagator. They become important for $E\sim M_s$ when
also stringy  effects become important. Note that this is much
earlier than the five-dimensional Planck scale. Moreover the bulk
$R^2$ term according to our analysis in section~\ref{sec:5},
becomes important at the compactification scale $M_c$ (this is
compatible with an effective brane thickness $w\sim M_s^{-1}$ at
energies below the winding threshold). However, the bulk $R^2$
terms in string theory are not expected to affect the graviton
propagator since they come in the Gauss-Bonnet combination.

In the case where all orbifold directions are non compact, our
analysis in section~\ref{sec:5} applies. Case (3) is the only one
where the interesting threshold can be smaller than the string
scale. Putting in numbers we obtain $L/r_c^2\sim (g_s
T_2)^{-1/2}M_s$, $1/r_c\sim (g_s^2 T_2)^{-1/4}M_s$ and $1/L\sim
g_s^{-1/2}M_s$. We are clearly in case (i) of
section~\ref{sec:5}. However all these scales are larger than the
compactification scale $M_c\sim M_s/\sqrt{T_2}$ so the
four-transverse dimensional discussion does not apply beyond
$M_c$. We learn though that gravity on the D4-brane is screened
up to the compactification scale, or for lengths larger than
$\sqrt{T_2}$. Beyond $M_c$ we have effectively a D4-brane in 10
non-compact dimensions. Thus, beyond $M_c$ gravity is  five
dimensional on the D4-brane.

\section{Conclusions}

The purpose of this paper was to provide a detailed analysis of
the behavior of theories with gravity induced on a brane through
quantum corrections by brane matter fields. We have relaxed
various approximations made in previous studies~\cite{dgp,dgkn}.

The first such approximation was the zero thickness of the brane.
We considered thick four-dimensional branes of thickness $\lb$ in
a  bulk. The natural choice is $\lb\sim 1/M$, with $M$ the
fundamental scale of the theory. We found that a new
characteristic scale emerges in this case: $E^{-1}_b=\sqrt{\lb
r_c}$, where $r_c\gg 1/M$ controls the strength of the induced
Einstein term on the brane. At distances larger than the new
scale the behavior in unaffected. However, at shorter distances
it deviates strongly from the expected four-dimensional behavior.
The solutions for the propagator and the KK modes vary strongly
within the brane, and we calculated their averages between the
edges. We found that the averaged gravitational potential displays
deviations from the four-dimensional form $\sim 1/r$ which depend
on the profile of the probe sources. This leads to violations of
the equivalence principle.

\looseness=-1
The main conclusion to be drawn from the above is that the
description in terms of the effective action loses its
predictability at energies much smaller than the fundamental
scale $M$. Only the knowledge of the structure of the fundamental
theory can predict correctly the behavior at energies larger than
the new threshold scale~$E_b$.

For the phenomenologically viable scenario in the case of a
compact fifth dimension the threshold is at experimentally
accessible distances of the submillimeter~range.

We also considered the next corrections in the derivative
expansion of the effective action: the $R^2$ terms in the bulk
and on the brane. The analysis for bulks with one or more extra
dimensions strengthens the conclusion reached above: \pagebreak[3] the
emergence of new threshold scales is generic. Several examples
were discussed in sections~\ref{sec:5} and~\ref{sec:6}. Moreover,
the new scales may appear in energy regimes where the effect of
the higher-derivative terms should be negligible according to
power-counting arguments. These results cast doubts on the quick
convergence of the derivative expansion in the context of the
induced gravity scenario.

However, for sufficiently small couplings the $R^2$ terms may
leave the basic structure of a model unaffected. For example, in
the case of one transverse dimension the induced ${\hat R}^2$
term on the brane does not change the large-distance behavior if
its coefficient $\lambda_2$ obeys $\lambda_2 \ll (M r_c)^3$. The
bulk $R^2$ term does not affect the large-distance physics and
becomes relevant only near the new threshold scale $E^{-1}_b$ we
discussed above. Thus the only modification to the model arises
at the experimentally accessible submillimeter range, where new
phenomena are expected.

We presented a string realization in which the $R$ and $R^2$
terms in the bulk and on the brane are calculable. New threshold
scales appear. The advantage of this model is that the physical
behavior on both sides of the new scales can be predicted with
confidence. In particular it can be seen that there are
transitions which imply that the five-dimensional Planck scale
does not correspond to a real threshold.

In our study of induced gravity when the transverse space has
more than one dimension we pointed out some interesting
transitions in the nature of gravity on the brane. For example,
in four large transverse dimensions without $R^2$ terms we found
that there is a screening of gravity at distances larger than
$r_0=\Lambda r_c^2$, with $\Lambda$ an ultraviolet cutoff. Such
screening could be allowed at cosmological distances beyond the
horizon. Taking $\Lambda \sim M$, $M_{Pl}^2 = M^6 r_c^4$ and $r_0
\sim 10^{26}\,\mathrm{m}$, we obtain $M\sim 5\times
10^{-3}\,\mathrm{eV}$ and $r_c\sim 10^{11}\,\mathrm{m}$. When an
$R^2$ correction is included in the bulk with $L/r^2_c \ll 1/r_c
\ll 1/L\ll \Lambda$ we found that we have screening of gravity at
distances larger than $r_0\sim r_c^2/L$. Here also we can fit the
Planck scale $M_{Pl}=M^3 r^2_c$ and choose $r_0$ to be larger
than the cosmological horizon, while keeping $L$ much smaller
than a millimeter. It remains to be seen whether realistic models
can be constructed along these lines.

The overall message of this work is that, in theories of induced
gravity on a brane, new phenomena appear at threshold scales well
below the higher-dimensional PLanck energy scale. The physics
above the new scales depends crucially on the structure of the
fundamental theory.

\acknowledgments

The work of N.~Tetradis was  partially supported through a RTN
contract HPRN--CT--2000--00148 of the European Union. The work of
E.~Kiritsis and T.~Tomaras was partially supported by RTN
contracts HPRN--CT--2000--00122 and --00131. We acknowledge also
partial support from INTAS grant N 99 1 590.

\end{document}